\documentstyle[epsbox,12pt]{article}
\newcommand{\simgt}{\lower.5ex\hbox{$\; \buildrel > \over \sim \;$}}
\newcommand{\simlt}{\lower.5ex\hbox{$\; \buildrel < \over \sim \;$}}
\begin{document}
\title{
%\vspace*{-1cm}
Effects of Neutrino Trapping
on Thermodynamic Properties of Nuclear ``Pasta''}
%Thermodynamic Properties of Nuclear ``Pasta'' in Supernova Cores}
\author{Gentaro Watanabe$^{\rm a}$, Kei Iida$^{\rm a,b}$, 
Katsuhiko Sato$^{\rm a,c}$\\
{\it $^{\rm a}$Department of Physics, University of Tokyo,
7-3-1 Hongo, Bunkyo,}\\ {\it Tokyo 113-0033, Japan}\\
{\it $^{\rm b}$Department of Physics, University of Illinois
at Urbana-Champaign,}\\ 
{\it 1110 West Green Street, Urbana, IL 61801-3080, USA}\\
{\it $^{\rm c}$Research Center for the Early Universe, 
University of Tokyo,}\\
{\it 7-3-1 Hongo, Bunkyo, Tokyo 113-0033, Japan}}
\maketitle

\baselineskip=24pt

%%%%%%%%%%%%%%%%%%%%%%
\begin{abstract}
%%%%%%%%%%%%%%%%%%%%%%

Geometrical structure of
matter at subnuclear densities is investigated in the presence
of a degenerate gas of neutrinos as encountered in stellar collapse.
The crystalline phases with spherical, cylindrical and planar nuclei 
as well as with spherical and cylindrical nuclear bubbles
are considered by using a compressible liquid-drop model.
This model allows for uncertainties
in the lepton fraction $Y_{\rm L}$ in addition to those in 
the nuclear surface tension $E_{\rm surf}$
and in the proton chemical potential in bulk neutron matter
$\mu_{\rm p}^{(0)}$.
The phase diagrams obtained at zero temperature show that
only the phases with rod-like and slab-like nuclei
occur at typical values of $Y_{\rm L}$,
$E_{\rm surf}$ and $\mu_{\rm p}^{(0)}$, whereas
the bubble phases, especially with spherical bubbles,
are at best expected at hypothetically low values of 
$Y_{\rm L}$ and/or $E_{\rm surf}$. 
For the rod-like and slab-like nuclei,
thermally induced displacements are calculated from
their respective elastic constants.
It is found that at temperatures appropriate to supernova cores,
thermal fluctuations would destroy 
the layered lattice of slab-like nuclei
almost independently of the nuclear models and of the degree of
the neutrino degeneracy.

\noindent
$PACS:$ 26.50.+x; 97.60.Bw

\noindent
$Keywords:$ Dense matter; Ground state; Thermal fluctuations; 
Stellar collapse

\end{abstract}

\newpage

%%%%%%%%%%%%%%%%%%%%%%%%%
\section{Introduction}
%%%%%%%%%%%%%%%%%%%%%%%%%

To clarify the mechanism of collapse-driven supernovae (see, e.g., 
Refs.\ \cite{bethe,suzuki} for review) necessitates 
understanding of the properties of
matter under extreme conditions characterized by
high densities and temperatures
as well as of its interactions with electron neutrinos,
which are produced by electron capture processes
on the nuclei present during the collapse of an iron core of a 
massive star.
Diffusion of the neutrinos thus produced is controlled by their
coherent scattering off the nuclei via neutral-current weak 
interactions \cite{freedman,sato}.
It is generally accepted that
at densities above $\sim10^{12}$ g cm$^{-3}$, 
where the dynamical time scale of the collapse is small compared 
with the diffusion time scale, these neutrinos are trapped,
forming an ultrarelativistic and degenerate Fermi gas. 
This plays a role in 
establishing $\beta$ equilibrium in the system\footnote{
Hereafter, the corresponding $\beta$ equilibrated material 
is referred to as supernova matter.}
and, until the core ceases to collapse and rebounds,
in increasing the system temperature $T$ up to order or greater than
10 MeV. 
The density region in which such a trapping occurs and  
the lepton fraction, i.e.,
the number of the leptons of electron flavor 
present per nucleon locally in this region:
\begin{equation}
Y_{\rm L} =\frac{n_{\rm e}+n_{\nu}}{n_{\rm b}}\ ,
\end{equation}
where $n_{\rm b}$, $n_{\rm e}$ and $n_{\nu}$ are the 
nucleon, electron and electron-neutrino number densities, 
respectively, depend on 
the rates of the electron capture and of the neutrino scattering.
These rates are in turn affected by changes in the states of 
supernova matter such as 
changes in the nuclear composition and a melting transition
from the phase with the nuclei to the phase of uniform nuclear matter.

At subnuclear densities and at temperatures below the critical 
temperature of order 10 MeV,
nuclear matter is separated into
a high-density liquid part, usually characterized by roughly spherical
nuclei, and a low-density gas part, composed mostly of neutrons. 
When the density is close to the saturation density of symmetric
nuclear matter, $\rho_{\rm s}\approx2.7\times10^{14}$ g cm$^{-3}$,
the system is expected to have rod-like and slab-like nuclei
as well as rod-like and roughly spherical bubbles,
as originally indicated by Ravenhall et al.\ \cite{ravenhall}
and Hashimoto et al.\ \cite{hashimoto} 
using a liquid drop model in the zero-temperature approximation.
These earliest investigations show that
such non-spherical nuclei and bubbles,
forming a Coulomb lattice,
arise from a delicate balance between the interfacial energy 
and the electrostatic energy including the lattice energy.
With increasing density, a bcc lattice of 
spherical nuclei changes into a two-dimensional triangular lattice
of rod-like nuclei and then into a one-dimensional layered lattice
of slab-like nuclei, until the liquid part and the gas part begin
to be replaced by each other.
After further transformations to a two-dimensional triangular lattice
of cylindrical bubbles and then to a bcc lattice of spherical 
bubbles, the system turns into uniform nuclear matter.
This behavior was confirmed by Lassaut et al.\ \cite{lassaut} 
in their three-dimensional Thomas-Fermi calculations 
at finite $Y_{L}$ and $T$ relevant to stellar collapse.
Those kinds of non-spherical nuclei and of bubbles are often 
referred to as nuclear ``pasta'' since they are similar in shape
to spaghetti, lasagna, etc.

The presence of non-spherical nuclei and of bubbles is
expected to have consequence to the hydrodynamics and 
neutrino transport in supernova cores; these are ingredients 
of a simulation study of stellar collapse, bounce and 
explosion (for recent progress in this study,
see, e.g., Ref.\ \cite{mezzacappa}).
Let us first note that
the neutrino wavelengths, typically of order 20 fm, 
are comparable to or even greater than the internuclear spacing,
leading to diffractive effects on the neutrino elastic scattering
off such a periodic spatial structure of nuclear matter 
\cite{ravenhall}.
These effects, induced by the internuclear Coulombic correlations,
would act to reduce the scattering rates and hence the lepton 
fraction $Y_{L}$.
For the bcc lattice of spherical nuclei, such a reduction was
examined by Horowitz \cite{horowitz} by calculating the 
associated static structure factor.
As clarified by Bruenn and Mezzacappa \cite{bruenn} in their 
numerical calculations of stellar collapse, however,
the resultant window that allows low-energy neutrinos to diffuse away
from the core during its collapse
is too narrow to drastically suppress the degree
of the neutrino degeneracy.
It is also noteworthy that 
non-spherical nuclei and bubbles are elongated in specific direction.
In such direction, the neutrino scattering processes are
no longer coherent, in contrast to the case of
roughly spherical nuclei whose finiteness in any direction
yields constructive interference in the scattering.
The last point to be mentioned is that
the changes in the nuclear shape are accompanied by
discontinuities in the adiabatic index, denoting how hard the
equation of state of the material is.
As Lassaut et al.\ \cite{lassaut} suggested, these discontinuities
may influence the core hydrodynamics during the initial phase of 
the collapse.

For neutron star matter, which has vanishing neutrino degeneracy,
we examined the density region 
in which the phases with non-spherical nuclei and with bubbles
are preferred over the bcc phase of spherical nuclei and 
the homogeneous phase 
by generalizing a compressible liquid-drop model
advanced by Baym, Bethe and Pethick \cite{BBP} (hereafter BBP)
so as to incorporate uncertainties in 
the nuclear surface tension $E_{\rm surf}$ and
the proton chemical potential $\mu_{\rm p}^{(0)}$
in the dripped neutron gas \cite{gentaro}.
We found that as $E_{\rm surf}$ decreases or $\mu_{\rm p}^{(0)}$
increases, such a density region is enlarged appreciably.
Fortunately, those uncertainties are of less importance 
to supernova matter
at subnuclear densities, which will be studied in the present paper.
This is because after the onset of the neutrino trapping, 
the resultant neutrino degeneracy keeps 
the denser part of nuclear matter
stable against further neutronization; not only does 
the proton fraction $x$ in this part
remain typically of order 0.3, not
far from the regime accessible to experiment, but also  
the dripped nucleon gas, arising mainly from thermal effects, 
is too low for the values of $\mu_{\rm p}^{(0)}$ to 
make any essential difference 
in the density region occupied by nuclear ``pasta''.
To be taken into account in analyzing supernova matter, in addition to 
the nuclear model dependence of the stability of the ``pasta'' phases,
is its dependence on the trapped lepton fraction $Y_{\rm L}$.
The values of $Y_{\rm L}$, depending on  
the sizes and shapes of the denser part which produces,
scatters and absorbs electron neutrinos via weak interactions, 
exert feedback on the condition for the stability of the
``pasta'' phases.
Note, for example, that with increasing $Y_{\rm L}$,
the melting density $\rho_{\rm m}$ approaches $\rho_{\rm s}$.
This is due to the tendency \cite{OTSST} that as the denser part
is at larger neutron excess, the saturation density is lowered.
By allowing for uncertainties in $Y_{\rm L}$ coming partly from 
such a feedback effect,
we set a range of $Y_{\rm L}$ as $0.2\leq Y_{\rm L}\leq0.4$,
rather wider than the usually adopted values ranging 0.3--0.38 
(see Refs.\ \cite{bethe,suzuki} and references therein).
The equilibrium phase diagrams of zero-temperature supernova matter
are drawn for this range of $Y_{\rm L}$; we thus find that
at typical values of $E_{\rm surf}$, $\mu_{\rm p}^{(0)}$ 
and $Y_{\rm L}$, the phases with rod-like and slab-like nuclei 
lie between the usual bcc phase and the uniform phase, 
as in the case of neutron star matter \cite{gentaro}.

For these two ``pasta'' phases, 
we evaluate thermally induced displacements of the nuclei
from their elastic constants, which were
described by Pethick and Potekhin \cite{potekhin}
in terms of the interfacial and electrostatic energies.
They noted that the low-dimensional spatial order exhibited by
the phases with cylindrical and planar nuclei 
is analogous to that of the liquid crystals, i.e.,
columnar phases and smectics A, respectively. 
According to
our previous analysis of neutron star matter \cite{gentaro},
thermal fluctuations
may melt the layered lattice of planar nuclei rather than
the triangular lattice of cylindrical nuclei
at temperatures appropriate to matter in neutron star crusts.
In the case of supernova cores of interest here, as we shall see,
not only the temperature scale but also the elastic constants
are considerably large as compared with 
the case of neutron star crusts.
The temperatures and elastic constants thus increased 
play a role in enhancing and 
suppressing the fluctuational displacements, respectively.
The resultant rivalry between these two roles is finally examined.

In Section 2, a compressible
liquid-drop model for nuclei and bubbles, 
together with the associated equilibrium conditions, 
is designed to describe zero-temperature matter 
at subnuclear densities and various lepton fractions.
In Section 3 we therefrom derive
the equilibrium phase diagrams for changes in the nuclear conformation.
Properties of such conformation changes are discussed in Section 4.
In Section 5 we estimate displacements of rod-like and slab-like nuclei 
at finite temperatures.
Section 6 is devoted to conclusions.

%%%%%%%%%%%%%%%%%%%%%%%%%%%%%%%%%%%%%%%%%%%%%%%%%%%%%%%%%%%%%%%%%%
\section{Model for supernova matter at subnuclear densities}
%%%%%%%%%%%%%%%%%%%%%%%%%%%%%%%%%%%%%%%%%%%%%%%%%%%%%%%%%%%%%%%%%%

In this section,
we construct the free energy of supernova matter for densities
near the melting density $\rho_{\rm m}$, and write down the
conditions for its equilibrium. 
We concentrate on the influence of the lepton fraction $Y_{\rm L}$
on the free energy.
Since the finite-temperature effects on the free energy, i.e.,
nuclear and electronic excitations, 
nucleon evaporation from
the denser part of nuclear matter,
multiple shapes and sizes of this part, etc.,
are of less importance, 
the temperature $T$ is taken to be zero.
All we bear in mind is thus to combine
the BBP-type compressible liquid-drop model
used by us \cite{gentaro} for the description of zero-temperature, 
inhomogeneous nuclear matter embedded in a sea of electrons
with the energy contribution of electron neutrinos.
At $T=0$, the less dense part of nuclear matter
is composed entirely of the neutrons dripped out of the
denser part.

Following Ref.\ \cite{gentaro},
we consider five phases characterized by the 
shapes of the denser part of nuclear matter, i.e., 
sphere, cylinder, slab, cylindrical hole and spherical hole,
respectively; each phase is composed of
a single species of nucleus or bubble at a given baryon density 
$n_{\rm b}$ and lepton fraction $Y_{\rm L}$.
The Wigner-Seitz approximation is adopted 
in describing the energy of the respective Coulomb lattice:
A cell in the bcc lattice, including
a spherical nucleus or bubble of radius $r_{\rm N}$, 
is replaced by a spherical Wigner-Seitz cell
having the same center and radius $r_{\rm c}$;
a cell in the two-dimensional triangular lattice, including  
a cylindrical nucleus or bubble having an infinitely long axis 
and a circular section of radius $r_{\rm N}$,
is replaced by a cylindrical Wigner-Seitz cell 
having the same axis and 
a circular section of radius $r_{\rm c}$;  
a cell in the one-dimensional layered lattice, including
a planar nucleus with width $2r_{\rm N}$, 
is equivalent to a planar Wigner-Seitz cell having the same
central plane and width $2r_{\rm c}$.
The values of $r_{\rm c}$ for these phases are chosen so that each
Wigner-Seitz cell may have zero net charge.

%%%%%%%%%%%%%%%%%%%%%%%%%%%%%%%%%%%
\subsection{Energy of matter}
%%%%%%%%%%%%%%%%%%%%%%%%%%%%%%%%%%%

Let us now regard each species of leptons (nucleons) 
as uniformly distributed everywhere in the system
(inside and outside the nuclei or bubbles), and
write down the total energy density $E_{\rm tot}$
of the system as 
\begin{equation}
E_{\rm tot}=\left\{
\begin{array}{ll}
w_{\rm N}+w_{\rm L}+(1-u)E_{\rm n}(n_{\rm n})
+E_{\rm e}(n_{\rm e})+E_{\nu}(n_{\nu})& \quad \mbox{(nuclei)}\ , \\ \\
w_{\rm N}+w_{\rm L}+uE_{\rm n}(n_{\rm n})
+E_{\rm e}(n_{\rm e})+E_{\nu}(n_{\nu})& \quad \mbox{(bubbles)}\ .
\end{array}\right.
\end{equation}
Here,
$w_{\rm N}$ is the energy of the nuclear matter region
(the region containing protons) 
in a cell as divided by the cell volume;
$w_{\rm L}$ is the lattice energy
per unit volume as given for 
the nuclei or bubbles having nonzero $r_{\rm N}$; 
$n_{\rm n}$ is the number density of neutrons outside the nuclei or 
inside the bubbles;
$E_{\rm n}$, $E_{\rm e}$ and $E_{\nu}$ are
the energy densities of the neutron matter, of the electron gas
and of the neutrino gas, respectively;
$u$ is the volume fraction occupied by the nuclei or bubbles.

In Eq.\ (2), the expressions for $w_{\rm N}$, $w_{\rm L}$, 
$E_{\rm n}$ and $E_{\rm e}$ are given 
by Eqs.\ (3), (8), (10) and (11) in Ref.\ \cite{gentaro}.  
A new component, the neutrino energy density, is expressed
in the form of an ideal ultrarelativistic Fermi gas:
\begin{equation}
E_{\nu}(n_{\nu})=\frac{3}{4}n_{\nu} \hbar k_{\nu}c\ ,
\end{equation}
where $k_{\nu}=(6\pi^2 n_{\nu})^{\frac{1}{3}}$ 
is the neutrino Fermi wave number.
Recall that the energy of the nuclear matter region
is written in the form of a compressible liquid-drop model
as (Eq.\ (3) in Ref.\ \cite{gentaro})
\begin{equation}
w_{\rm N}(n,x,n_{\rm n},r_{\rm N},r_{\rm c},d) = \left\{
\begin{array}{ll}
un[(1-x)m_{\rm n}+xm_{\rm p}]c^2+unW(k,x)\nonumber\\
\quad
+w_{\rm surf}(n,x,n_{\rm n},r_{\rm N},u,d)
+w_{\rm C}(n,x,r_{\rm N},u,d)
& \quad \mbox{(nuclei)\ ,}\\ \\
(1-u)n[(1-x)m_{\rm n}+xm_{\rm p}]c^2+(1-u)nW(k,x)\nonumber\\
\quad
+w_{\rm surf}(n,x,n_{\rm n},r_{\rm N},u,d)
+w_{\rm C}(n,x,r_{\rm N},u,d)
& \quad \mbox{(bubbles)\ ,}
\end{array}\right.
\end{equation}
where $m_{\rm n}$ $(m_{\rm p})$ is the neutron (proton) rest mass;
$n$ is the nucleon number density in the nuclear matter region;
$W(k,x)$ is the energy per nucleon for uniform nuclear matter of nucleon 
Fermi wave number $k=(3\pi^2 n/2)^{1/3}$ and proton fraction $x$;
$w_{\rm surf}$ is the nuclear surface energy per unit volume;
$w_{\rm C}$ is the self Coulomb energy (per unit volume)
of protons contained in a cell;
$d$ is the dimensionality defined as $d=1$ for slabs,
$d=2$ for cylinders and $d=3$ for spheres.
This expression for $w_{\rm N}$ includes
three parameters $C_{1}$, $C_{2}$ and $C_{3}$, associated with
uncertainties in the proton chemical potential $\mu_{\rm p}^{(0)}$ 
in pure neutron matter as contained in $W(k,x)$ 
and those in the nuclear surface tension
$E_{\rm surf}(=r_{\rm N}w_{\rm surf}/ud)$. 
$C_{1}$ determines the magnitude of $\mu_{\rm p}^{(0)}$
(not including the rest mass)
as (Eq.\ (4) in Ref.\ \cite{gentaro})
\begin{equation}
\mu_{\rm p}^{(0)}=-C_1 n_{\rm n}^{2/3}\ .
\end{equation} 
We basically set $C_{1}=400$ MeV fm$^{2}$; this case is consistent 
with the generally accepted behavior among recent literature 
as exhibited in Fig.\ 1 of Ref.\ \cite{gentaro}.
As the case may be,
$C_{1}=300, 500, 600$ MeV fm$^{2}$ will be also taken
for the sake of comparison.
The other two parameters $C_{2}$ and $C_{3}$ are 
defined as (Eq.\ (6) in \cite{gentaro})
\begin{equation}
E_{\rm surf}=C_{2}\tanh\left(\frac{C_{3}}{\mu_{\rm n}^{(0)}}\right) 
E_{\rm surf}^{\rm BBP}\ ,
\end{equation}
where 
$\mu_{\rm n}^{(0)}=\partial E_{\rm n}/\partial n_{\rm n} - m_{\rm n}c^2$ 
is the neutron chemical potential in the neutron gas
(not including the rest mass), and $E_{\rm surf}^{\rm BBP}$
is the BBP-type surface tension
(Eq.\ (7) in Ref.\ \cite{gentaro}).
As shown in Ref.\ \cite{gentaro}
for the ground-state matter with $n_{\nu}=0$ and 
$C_{1}=300, 400, 500, 600$ MeV fm$^{2}$,
to set $C_{2}=1.0$ and $C_{3}=3.5$ MeV 
well reproduces the Hartree-Fock (HF) results 
obtained by Ravenhall, Bennett and Pethick \cite{RBP} (hereafter
denoted by RBP) using a Skyrme interaction.
It is important to examine whether or not this tendency
persists for the ground-state supernova matter;
in this case, the proton fraction $x$ of interest
is considerably large relative to the case of $n_{\nu}=0$ 
in which it is confined to $x<0.2$.
When we still set $C_{2}=1.0$ and  $C_{3}=3.5$ MeV, 
as can be seen from Fig.\ 1,
the surface tension calculated 
from the equilibrium conditions to be described in Subsection 2.2
agrees fairly well with the RBP results
for $C_{1}=400$ MeV fm$^{2}$ and $Y_{\rm L}=0.2,0.3,0.4$.\footnote
{At fixed $Y_{\rm L}$ and for densities of interest here,
the proton fraction $x$ is limited to a rather small range 
exhibited in Fig.\ 1.} 
Such agreement has been confirmed also for
$C_{1}=300, 500, 600$ MeV fm$^{2}$ and $Y_{\rm L}=0.2,0.3,0.4$.
In the present work, we thus give $C_{3}$ a fixed value, 3.5 MeV,  
and $C_{2}$ a range of values including unity, 0.01--3.

%%%%%%%%%%%%%%%%%%%%%%%%%%%%%%%%%%%%%%%%%
\subsection{Equilibrium conditions}
%%%%%%%%%%%%%%%%%%%%%%%%%%%%%%%%%%%%%%%%%

Zero-temperature supernova matter with nuclei or bubbles of given shape,
in its equilibrium,
fulfills the conditions for stability of the nuclear matter
region against change in the size, neutron drip, $\beta$-decay and 
pressurization, 
as in the case of neutron star matter (see Section 2.2 in Ref.\ 
\cite{gentaro}).
These conditions come from minimization of 
the energy density $E_{\rm tot}$
with respect to five variables $n$, $x$, $n_{\rm n}$, $r_{\rm N}$
and $u$ at fixed lepton fraction $Y_{\rm L}$ given by Eq.\ (1)
and baryon density $n_{\rm b}$ given by
\begin{equation}
n_{\rm b} = \left\{
\begin{array}{ll}
un+(1-u)n_{\rm n} & \quad \mbox{(nuclei)\ ,}\\
(1-u)n+un_{\rm n} & \quad \mbox{(bubbles)\ ,}
\end{array}\right.
\end{equation}
as well as under charge neutrality,
\begin{equation}
n_{\rm e} = \left\{
\begin{array}{ll}
xnu & \quad \mbox{(nuclei)\ ,}\\
xn(1-u) & \quad \mbox{(bubbles)\ .}
\end{array}\right.
\end{equation}
The resultant expressions for
the size, drip and pressure equilibria are the same as
those given for neutron star matter 
(see Eqs.\ (14), (15) and (21) in Ref.\ \cite{gentaro}).
Only the $\beta$-equilibrium condition 
(Eq.\ (18) in Ref.\ \cite{gentaro}) is modified
by the neutrino degeneracy; it now reads
\begin{equation}
\mu_{\rm e}-\mu_{\nu}-(m_{\rm n}-m_{\rm p})c^{2}=
\mu_{\rm n}^{\rm (N)}-\mu_{\rm p}^{\rm (N)}\ ,
\end{equation}
where
\begin{equation}
\mu_{\nu}=\hbar k_{\nu} c
\end{equation}
is the neutrino chemical potential; $\mu_{\rm e}$, 
$\mu_{\rm n}^{\rm (N)}$ and $\mu_{\rm p}^{\rm (N)}$ 
are the electron chemical potential, the neutron chemical potential
in the nuclear matter region and the proton chemical
potential in the nuclear matter region, respectively
(see Eqs.\ (19), (16) and (20) in Ref.\ \cite{gentaro}).

At given $n_{\rm b}$, $Y_{\rm L}$ and nuclear shape,
we have calculated the minimum value of $E_{\rm tot}$ 
for electrically neutral supernova matter.
Such calculations begin with
elimination of $\mu_{\nu}$ from condition (9)
with the help of Eqs.\ (1) and (10),
leading to a cubic equation for $\mu_{\rm e}^{\frac{1}{3}}$.
The solution for $\mu_{\rm e}$ 
can then be written in terms of $n$, $x$, $n_{\rm n}$ and $r_{\rm N}$
by using the Cardan's formula.
For the rest of the process that allows us to 
find the equilibrium values of 
$n$, $x$, $n_{\rm n}$, $r_{\rm N}$ and $r_{\rm c}$
and thus the minimum value of $E_{\rm tot}$,
we can follow a line of argument of Ref.\ \cite{gentaro}.

At the same $n_{\rm b}$ and $Y_{\rm L}$,
the energy density of
uniform nuclear matter in $\beta$-equilibrium with leptons
has also been evaluated. 
The total energy density of this uniform system
is written as
\begin{equation}
E_{\rm tot} = nW(k,x)+n[(1-x)m_{\rm n}+xm_{\rm p}]c^2
+E_{\rm e}(n_{\rm e})+E_{\nu}(n_{\nu})\ ,
\end{equation}
where the presence of muons is ignored.
We then minimize $E_{\rm tot}$ with respect to $x$
at fixed baryon density $n_{\rm b}(=n)$ and lepton fraction $Y_{\rm L}$
as well as subject to charge neutrality,
\begin{equation}
n_{\rm e}=xn\ .
\end{equation}
The resultant $\beta$-equilibrium condition, given by
\begin{equation}
\frac{\partial W(k,x)}{\partial x} =
-\mu_{\rm e}+\mu_{\nu}+(m_{\rm n}-m_{\rm p})c^2\ ,
\end{equation}
leads to the optimal value of $x$ and hence of $E_{\rm tot}$. 
By comparing this value of $E_{\rm tot}$ with the values obtained 
for the five crystalline phases, 
we have determined the phase giving the smallest energy density
at various values of $n_{\rm b}$, $Y_{\rm L}$, $C_1$ and $C_2$,
and thereby drawn the phase diagrams for the ground-state
supernova matter, as plotted in Figs.\ 2 and 3.

%%%%%%%%%%%%%%%%%%%%%%%%%%%%%%%%%
\section{Phase diagrams}
%%%%%%%%%%%%%%%%%%%%%%%%%%%%%%%%%

Let us first examine the $\mu_{\rm p}^{(0)}$ and $E_{\rm surf}$ 
dependence of the density region of each phase;
such dependence can be typically seen from Fig.\ 2 that exhibits
the phase diagram drawn at $Y_{\rm L}=0.3$ and 
$C_1=300, 400, 600$ MeV fm$^2$
over the $n_{\rm b}$ versus $C_{2}$ plane.
We have thus confirmed the feature as mentioned in Section 1 that 
the density region where neither the usual bcc phase nor the uniform phase
is stable is almost independent of $\mu_{\rm p}^{(0)}$
because of negligibly small $n_{\rm n}$.
We remark in passing that
the melting density $\rho_{\rm m}$
increases with increasing $C_1$ (lowering $\mu_{\rm p}^{(0)}$),
in contrast to the case of neutron star matter
(see Fig.\ 3 in Ref.\ \cite{gentaro}).
This is because such a change in $\mu_{\rm p}^{(0)}$
helps the nuclear matter region keep the protons within itself,
rather than the neutron gas region share the protons. 
The $E_{\rm surf}$ dependence, on the other hand, is essentially
the same as that obtained for neutron star matter:
Not only is the density region occupied by nuclear ``pasta''
reduced with increasing $C_{2}$, but also 
the sequence of the nuclear shape for the lower $E_{\rm surf}$
($C_2\simlt0.3$), i.e.,  
sphere $\rightarrow$
cylinder $\rightarrow$ slab $\rightarrow$ cylindrical hole
$\rightarrow$ spherical hole $\rightarrow$ uniform matter
(with increasing $n_{b}$), changes into
sphere $\rightarrow$ cylinder $\rightarrow$ slab 
$\rightarrow$ uniform matter
for the typical $E_{\rm surf}$ ($C_2\simeq1.0$).  
However, it is to be recalled that uncertainties 
accompanying $C_{2}$ are smaller for supernova matter 
than for neutron star matter (see Section 1).

We proceed to consider the $Y_{\rm L}$ dependence of the 
phase diagrams by observing Fig.\ 3 drawn
for $Y_{\rm L}=0.2, 0.25, 0.3, 0.4$ and for $C_1=400$ MeV fm$^2$.
For these values of $Y_{\rm L}$, 
within the typical $E_{\rm surf}$ region ($C_{2}\simeq 1.0$),
the phase with slab-like nuclei dissolves into uniform matter
without undergoing a transition to the phase with
cylindrical or spherical bubbles. 
The most viable situation in which 
at least one of the bubble phases is energetically favored
for a finite range of density is expected  
in a rather hypothetical parameter region, $Y_{\rm L}\simeq0.2$ 
and $C_2\simeq0.7$.
It is interesting to note that as $Y_{\rm L}$ increases,
the melting density $\rho_{\rm m}$ goes up and becomes close to 
$\rho_{\rm s}$.
As mentioned in Section 1, this is due to the proton fraction 
dependence of the saturation density; for details, 
see Fig.\ 1 in Ref.\ \cite{arponen} that shows  
the BBP model for the energy of bulk nuclear matter 
underlying $W(k,x)$ used here.
We also find that the density region occupied by nuclear ``pasta''
is considerably larger than that displayed by Fig.\ 3 
in Ref.\ \cite{gentaro}
for neutron star matter having a lepton fraction of order 0.03
and a surface tension of order 0.1 MeV fm$^{-2}$.
This is because
such a density region is controlled by the tendency towards 
reduction in the total surface area, which 
is in turn determined by $E_{\rm surf}$
and hence by $Y_{\rm L}$ (see Fig.\ 1).
This tendency communicates with the fact that for supernova matter,
the energy difference between two successive phases is 
generally of order 1-10 keV per nucleon (see Fig.\ 4), 
an order of magnitude larger than that for neutron star matter 
(see Fig.\ 4 in Ref.\ \cite{gentaro}).

We conclude this section by asking why the bubble phases 
do not appear in the equilibrium phase diagrams
for our typical nuclear model ($C_1=400$ MeV fm$^2$ and $C_{2}=1.0$)
in contrast with the results obtained by Lassaut et al.\ 
\cite{lassaut} in their three-dimensional Thomas-Fermi calculations 
using a Skyrme nucleon-nucleon interaction. 
As expected from a simple liquid-drop argument of the electrostatic 
and interfacial effects ignoring the bulk energy \cite{hashimoto},
they found that
all the phases with non-spherical nuclei and with bubbles 
occur for a finite range of pressure (or density).
We ascribe this contrast primarily to the difference in the 
adopted properties of bulk nuclear matter at a proton fraction 
of about 0.3; 
description of these properties relies on interpolation 
between the fairly-well-determined properties of pure 
neutron matter and symmetric nuclear matter. 
This ascription is supported by the speculation
that the role of 
curvature corrections in stabilizing the phases with bubbles 
relative to the phases with nuclei,
which was allowed for in Ref.\ \cite{lassaut} 
but has been disregarded in the present work, 
is not sufficient to cause the bubble phases to appear
in the vicinity of $C_2=1.0$ on the phase diagrams. 
Such insufficiency can be viewed from Fig.\ 5, in which
the curvature energy gain per nucleon
for spherical bubbles over for planar nuclei
and the difference in energy per nucleon between
the phases with spherical bubbles and with planar nuclei
have been compared for $Y_{\rm L}=0.3$,
$C_1=400$ MeV fm$^2$ and $C_{2}=1.0$.\footnote{
Note that for this parameter set and at densities 
close to the melting point $\rho_{\rm m}$, the phase with 
cylindrical bubbles is less favorable
than that with spherical bubbles
even in the absence of the curvature corrections which
lower the energy for the spherical-bubble phase further than
that for the cylindrical-bubble phase.}
The curvature energy for given conformation 
is expressed per nucleon as 
\begin{equation}
W_{\rm curv}=
\left\{
\begin{array}{ll}
\displaystyle{\frac{d(d-1)u\omega_{\rm c}}{n_{\rm b}r_{\rm N}^{2}}} 
& \quad \mbox{(nuclei)\ ,}\\
-\displaystyle{\frac{d(d-1)u\omega_{\rm c}}{n_{\rm b}r_{\rm N}^{2}}} 
& \quad \mbox{(bubbles)\ ,}
\end{array}\right.
\end{equation} 
where $\omega_{\rm c}$ is
the curvature thermodynamic potential per unit length. 
We have set the values of $\omega_{\rm c}$, which have
yet to be determined well even at proton fractions near $x=0.5$, 
as $\omega_{\rm c}=0.1,0.2,0.5$ MeV fm$^{-1}$ by reference to 
the estimates, $\sim0.1$--0.2 MeV fm$^{-1}$ at $x\sim0.3$,
made by Kolehmainen et al.\ \cite{kolehmainen} 
in the Thomas-Fermi theory with Skyrme interactions.
Figure 5 indicates 
that for the typical parameter set leading to $x\sim0.3$,
the value of $\omega_{\rm c}$
required by the presence of the bubble phases is
at least $\sim0.5$ MeV fm$^{-1}$, considerably 
larger than the estimates referred to.
For $\omega_{\rm c}\sim0.1$--0.2 MeV fm$^{-1}$,
all the curvature corrections contribute to the vicinity of
$C_2=1.0$ on the phase diagrams
is to lower the transition density
from sphere to cylinder and that from cylinder to slab
by several tens percent and to accordingly enlarge
the density region surrounded by the bcc phase
and the uniform phase.

%%%%%%%%%%%%%%%%%%%%%%%%%%%%%%%%
\section{Properties of phase transitions}
%%%%%%%%%%%%%%%%%%%%%%%%%%%%%%%%

At a pressure at which 
the transition between one nuclear shape and another occurs, 
various quantities such as $n_{\rm b}$, $r_{\rm c}$, $r_{\rm N}$,
$n$, $x$ and $n_{\rm n}$ are more or less discontinuous
because of the dependence of the 
interfacial and electrostatic energies on the 
dimensionality $d$ (see Eqs.\ (5) and (8) in Ref.\ \cite{gentaro}).
As can be derived from Fig.\ 4 by implementing 
the double-tangent constructions 
denoting the coexistence of the two neighbouring phases,
the discontinuity in $n_{\rm b}$ amounts to of order 1\%.
Due to such a weak first-order nature of the transition, 
the transition point, properly characterized by a constant pressure,
is well described by a constant density determined by the 
comparison between the total energy densities for 
the respective phases.

Let us then examine the changes in 
the sizes of the nucleus or bubble
and of the Wigner-Seitz cell,
$r_{\rm N}$ and $r_{\rm c}$,
associated with the structural transitions.
In Fig.\ 6 we have plotted the values of 
$r_{\rm N}$ and $r_{\rm c}$
evaluated as functions of $n_{\rm b}$ for $Y_{\rm L}=0.3$,
$C_1=400$ MeV fm$^2$ and $C_2=0.01, 1.0, 2.5$.
Discontinuities in $r_{\rm N}$ and $r_{\rm c}$ 
at the transition points have been thus clarified.
In the case of $C_2=0.01$ in which all the five crystalline
phases appear, we observe  
the same dimensionality dependence of $r_{\rm c}$ 
as observed for neutron star matter 
(see Fig.\ 5 in Ref.\ \cite{gentaro}),
i.e., it is largest for $d=3$ (spheres and cylindrical holes)
and smallest for $d=1$ (slabs).

We finally address the question of 
how the neutron densities
outside and inside the nuclear matter region change
as the system melts into uniform nuclear matter.
Comparison of these densities with the one for the uniform phase
has been made in Fig.\ 7 
for $Y_{\rm L}$=0.3, $C_1$=400 MeV fm$^2$ and 
$C_2=0.01, 0.1, 1.0, 2.5$.
We can observe from this figure that for densities up to
the melting point, 
the density of the dripped neutron gas remains negligibly small
compared with the neutron density inside the nuclear matter region.
This behavior, which
stems from the neutrino trapping (see Section 1) and 
manifests itself for the other values of $Y_{\rm L}$ and $C_1$,
is consistent with the results 
obtained by Ogasawara and Sato \cite{ogasawara}
in their Thomas-Fermi calculations at various values of  
$Y_{\rm L}$ and zero temperature.
We also find from the comparison of Fig.\ 7 with 
Fig.\ 6 in Ref.\ \cite{gentaro}
that the difference in the neutron density profiles
between the crystalline and uniform phases 
at the melting point is remarkable in contrast with the case of
neutron star matter, 
in which the neutron density profile is fairly levelled
just below the melting point.
In supernova matter, in which
neighbouring nuclei (or bubbles) appear to touch and fuse each other 
at the melting point,
neutrino elastic scattering off nuclear ``pasta''
is thus expected to take effect until it becomes uniform.

%%%%%%%%%%%%%%%%%%%%%%%%%%%%%%%%%%%%%%%
\section{Thermal fluctuations}
%%%%%%%%%%%%%%%%%%%%%%%%%%%%%%%%%%%%%%%

Given occurrence of the phases with rod-like and slab-like nuclei
as expected for typical values of $E_{\rm surf}$ and $Y_{\rm L}$ from 
the zero-temperature phase diagrams exhibited in the previous section,
we turn to the estimates of thermally induced displacements of 
these nuclei from their equilibrium positions.
In making such estimates,  
we utilize the expressions for such displacements
that were written in Ref.\ \cite{gentaro} on the basis of
the elastic constants \cite{potekhin} controlling 
small deformation of the nuclei.
For the layered lattice of slab-like nuclei, 
the mean square of 
the displacement $v$ of layers in their normal direction reads
(Eq.\ (36) in Ref.\ \cite{gentaro})
\begin{equation}
\langle |v|^2 \rangle \simeq \frac{k_{\rm B} T}{4\pi \sqrt{BK_1}}
\ln{\left( \frac{L}{a} \right)}\ ,
\end{equation}
where $k_{B}$ is the Boltzmann constant,
$a=2r_{\rm c}$ is the layer spacing, 
$B$ and $K_1$ are the elastic constants given by Eqs.\ (28) and (29)
in Ref.\ \cite{gentaro}, respectively, and
the length scale $L$ of the lattice is assumed to be far larger 
than $r_{\rm c}$.
For the two-dimensional triangular lattice of rod-like nuclei, 
the mean square of 
the displacement vector $\mbox{\boldmath $v$}$
of cylinders in a plane perpendicular
to their elongated direction 
is expressed as (Eq.\ (37) in Ref.\ \cite{gentaro})
\begin{equation}
\langle 
|\mbox{\boldmath $v$}|^{2}
\rangle
\simeq \frac{k_{\rm B} T}{(B+2C) \sqrt{\pi\lambda a}}\ ,
\end{equation}
where $\lambda = \sqrt{2K_3/(B+2C)}$,
$B$, $C$ and $K_3$ are the elastic constants 
given by Eqs.\ (31), (34) and (35)
in Ref.\ \cite{gentaro}, respectively, 
$a=(2\pi /\sqrt{3})^{1/2} r_{\rm c}$ is the lattice constant
for hexagonal cells, and
the linear dimension of the lattice
in the plane including $\mbox{\boldmath $v$}$ is assumed to be
much larger than $r_{\rm c}$ and to be much smaller than
the length of the nuclei.
It is important to notice that 
all the elastic constants, derived 
by Pethick and Potekhin \cite{potekhin} from the calculations of the
electrostatic and interfacial energy increase
due to the nuclear deformation,
are proportional to the electrostatic energy density,
$w_{\rm C+L}$, given by Eq.\ (8) in Ref.\ \cite{gentaro}.
Since $w_{\rm C+L}$ in turn behaves as $\propto x^{2}$, 
the non-spherical nuclei in supernova cores
are more difficult to deform than those in neutron star crusts.
As was noted in \cite{gentaro} concerning
the crossover temperature $T_{0}$ from the quantum-fluctuation
to the thermal-fluctuation regime,
expressions (15) and (16) hold for $T\simgt T_{0}$
but underestimate the displacements for $T\simlt T_{0}$.
In the case of supernova matter considered here,
$T_{0}$ can be roughly estimated to be 
of order $10^{7}$ K for planar nuclei and 
of order $10^{9}$ K for cylindrical nuclei.

Let us then ask how large the root-mean-square displacement
$\sqrt{\langle |\mbox{\boldmath $v$}|^{2} \rangle}$
of a planar or cylindrical nucleus is relative to the shortest 
distance, $a/2 - r_{\rm N}$, between the surface of the nucleus
in its equilibrium position and
the boundary of the cell containing it.
The ratio of these two lengths for the planar (cylindrical) nucleus
has been plotted in Fig.\ 8 (9) at $T=10$ MeV and $L=10$ km,
appropriate to the subnuclear density regime inside supernova cores;
the $L$ dependence of the displacement, which is logarithmic,
makes little differences. 
We can find from these figures
the dominance of the relative displacement of  
the planar nucleus over that of the cylindrical nucleus,
yielded by the logarithmic factor appearing in Eq.\ (15),
as well as
the suppression of the relative displacements by $E_{\rm surf}$,
which were obtained also for neutron star crusts (see Figs.\ 7 and 8 
in Ref.\ \cite{gentaro}).
This $E_{\rm surf}$ dependence then communicates with
the suppression of the relative displacements by $Y_{L}$ 
as exhibited in Figs.\ 8 and 9 via the relation
between  $E_{\rm surf}$ and $Y_{L}$ shown in Fig.\ 1.
In examining the competition between 
the influences of temperature and of elasticity
as pointed out in Section 1,
it is noteworthy that the relative displacements thus obtained
are larger than those obtained for neutron star crusts
by less than a factor of ten.
This is because the influence of 
the difference in the typical temperature 
by two orders-of-magnitude
surpasses that of the above-mentioned difference in the
elasticity of the nucleus.
In case we set the melting condition as 
$\sqrt{\langle |\mbox{\boldmath $v$}|^{2} \rangle}=a/2 - r_{\rm N}$,
the enhanced relative displacements in supernova cores
do not modify our previous conclusion \cite{gentaro}
obtained for neutron star crusts
that the triangular lattice of rod-like nuclei is stable
against thermal fluctuations for our typical nuclear model,
but change the conclusion about the 
layered lattice of slab-like nuclei
from its marginal instability \cite{gentaro} depending on the 
nuclear models to its instability almost independent of them.
Given uncertainties in the temperature profile of the core,
it is useful to estimate the critical temperature $T_{\rm c}$
at which the relative displacements become unity; 
the results for $T_{\rm c}$ have been plotted 
in Figs.\ 10 and 11 for the layered lattice and 
the triangular lattice, respectively.

%%%%%%%%%%%%%%%%%%%%%%%%%%
\section{Conclusion}
%%%%%%%%%%%%%%%%%%%%%%%%%%

The dependence on the lepton fraction $Y_{\rm L}$ of the
phase diagrams associated with the changes in the nuclear shape
has been studied for zero-temperature
supernova matter at subnuclear densities.
We have found that with increasing $Y_{\rm L}$,
the density region occupied by the ``pasta'' phases, i.e.,
the phases with non-spherical nuclei and with bubbles, 
becomes larger.
Despite the considerable increase in the lepton fraction 
from neutron star matter, leading to enhancement of  
the proton fraction in the nuclear matter region and 
suppression of the density of the dripped neutron gas, not only
the sequence of the nuclear shape (sphere $\rightarrow$
cylinder $\rightarrow$ slab $\rightarrow$ uniform
with increasing density) obtained for the typical nuclear model
but also the effectiveness of thermal fluctuations
at the dissolution of slab-like nuclei
has been shown to be analogous to the case of neutron star matter.

The actual presence of nuclear ``pasta'' in supernova cores
depends on how the uncertain properties of bulk nuclear
matter having a proton fraction of order 0.3 
affect the stability of the ``pasta''
and on whether or not the time needed to form the ``pasta'' 
is small compared with the time scale of stellar collapse.
If future work about these questions
were in favor of the occurrence of the ``pasta'',
it would be significant to calculate
the finite-temperature equation of state and
neutrino transport in the presence of the ``pasta'' phases 
and to incorporate them into a supernova simulation 
with a neutrino radiation hydrodynamics code.
This would clarify the influence of the ``pasta''
on the lepton fraction at the onset of the neutrino trapping.

%%%%%%%%%%%%%%%%%%%%%%%%%%%%%%%%%
\section*{Acknowledgements}
%%%%%%%%%%%%%%%%%%%%%%%%%%%%%%%%%

This work was supported in part by 
Grants-in-Aid for Scientific Research provided by the Ministry of
Education, Science and Culture of Japan through Research Grant
No.\ 07CE2002 and No.\ 10-03687.

\newpage

Fig.\ 1. The surface energy per unit area (the surface tension)
as a function of $x$.
The thick lines are the present results obtained from Eq.\ (6)
for $C_{1}=400$ MeV fm$^{2}$, $C_{2}=0.01, 0.1, 0.5, 1.0, 2.5$ and
$Y_{\rm L}=0.2, 0.3, 0.4$.
The solid curve is the RBP result from their HF calculation
\cite{RBP}, and the dotted curve is the BBP result \cite{BBP}.
The cross denotes the value adopted by Ravenhall et al.\ \cite{ravenhall}.

Fig.\ 2. Zero-temperature phase diagrams
on the $n_{\rm b}$ versus $C_2$ plane,
evaluated for $Y_{\rm L}=0.3$ and $C_1=300, 400, 600$ MeV fm$^{2}$.

Fig.\ 3. Zero-temperature phase diagrams
on the $n_{\rm b}$ versus $C_2$ plane,
evaluated for $Y_{\rm L}=0.2, 0.25, 0.3, 0.4$ and $C_1=400$ MeV fm$^{2}$.

Fig.\ 4. Energy per nucleon relative to that of uniform matter
calculated for $Y_{\rm L}=0.3$, $C_1=400$ MeV fm$^2$ and $C_2=1.0$
as a function of baryon density $n_{\rm b}$.
The symbols SP, C, S, CH and SPH stand for sphere, cylinder, slab,
cylindrical hole and spherical hole, respectively.

Fig.\ 5. Energy per nucleon of the phase with spherical bubbles
relative to that with slab-like nuclei,
calculated for $Y_{\rm L}=0.3$, $C_1=400$ MeV fm$^2$ and $C_2=1.0$
as a function of baryon density $n_{\rm b}$.
The absolute value of
the curvature energy per nucleon of the phase with spherical bubbles
is also evaluated for the curvature thermodynamic potential per unit length
$\omega_{\rm c}=0.1,0.2,0.5$ MeV fm$^{-1}$.

Fig.\ 6. Size of a nucleus or bubble, $r_{\rm N}$, and 
of a Wigner-Seitz cell, $r_{\rm c}$,
calculated for $Y_{\rm L}=0.3$, $C_{1}=400$ MeV fm$^{2}$
and $C_{2}=0.01, 1.0, 2.5$ 
as a function of baryon density $n_{\rm b}$.
The symbols SP, C, S, CH and SPH stand for 
sphere, cylinder, slab, cylindrical hole and spherical hole, 
respectively.

Fig.\ 7. The neutron densities obtained
for $Y_{\rm L}=0.3$, $C_{1}=400$ MeV fm$^{2}$
and $C_{2}=0.01, 0.1, 1.0, 2.5$ as a function of baryon density $n_{\rm b}$.
The lines classified by $[n(1-x)]_{\rm N}$ and $n_{\rm n}$
represent the neutron densities in the nuclear matter region 
and in the neutron gas region for the phases with nuclei, respectively.
The lines classified by $[n(1-x)]_{\rm uni}$ denote the neutron densities
in uniform nuclear matter.

Fig.\ 8. The root-mean-square displacement of a planar nucleus
at $k_{\rm B}T=10$ MeV,
divided by the shortest distance 
between the surface of the nucleus in its equilibrium 
position and the boundary of the cell containing it.
The curves are obtained for $C_{1}=400$ MeV fm$^2$ and
for various sets of $Y_{\rm L}$ and $C_{2}$
as a function of baryon density $n_{\rm b}$.
The thick curves lying between the two vertical lines are the results 
in the density region in which the phase with planar nuclei is 
energetically stable.

Fig.\ 9. The root-mean-square displacement of a cylindrical nucleus
at $k_{\rm B}T=10$ MeV,
divided by the distance between 
the surface of the nucleus in its equilibrium position 
and the boundary of the cell containing it.
The curves are obtained for $C_{1}=400$ MeV fm$^2$ and
for various sets of $Y_{\rm L}$ and $C_{2}$
as a function of baryon density $n_{\rm b}$.
The thick curves lying between the two vertical lines are the results 
in the density region in which the phase with cylindrical nuclei is 
energetically stable.

Fig.\ 10. The critical temperature $T_{\rm c}$
for the phase with planar nuclei
as a function of baryon density $n_{\rm b}$.
The thick curves lying between the two vertical lines are the results 
in the density region in which the phase with planar nuclei is
energetically stable.
$C_1$ is taken to be 400 MeV fm$^2$.

Fig.\ 11. The critical temperature $T_{\rm c}$
for the phase with cylindrical nuclei
as a function of baryon density $n_{\rm b}$.
The thick curves lying between the two vertical lines are the results 
in the density region in which the phase with cylindrical nuclei is
energetically stable. 
$C_1$ is taken to be 400 MeV fm$^2$.

\newpage
\begin{figure}
  \begin{center}
  \psbox[height=15.5cm]{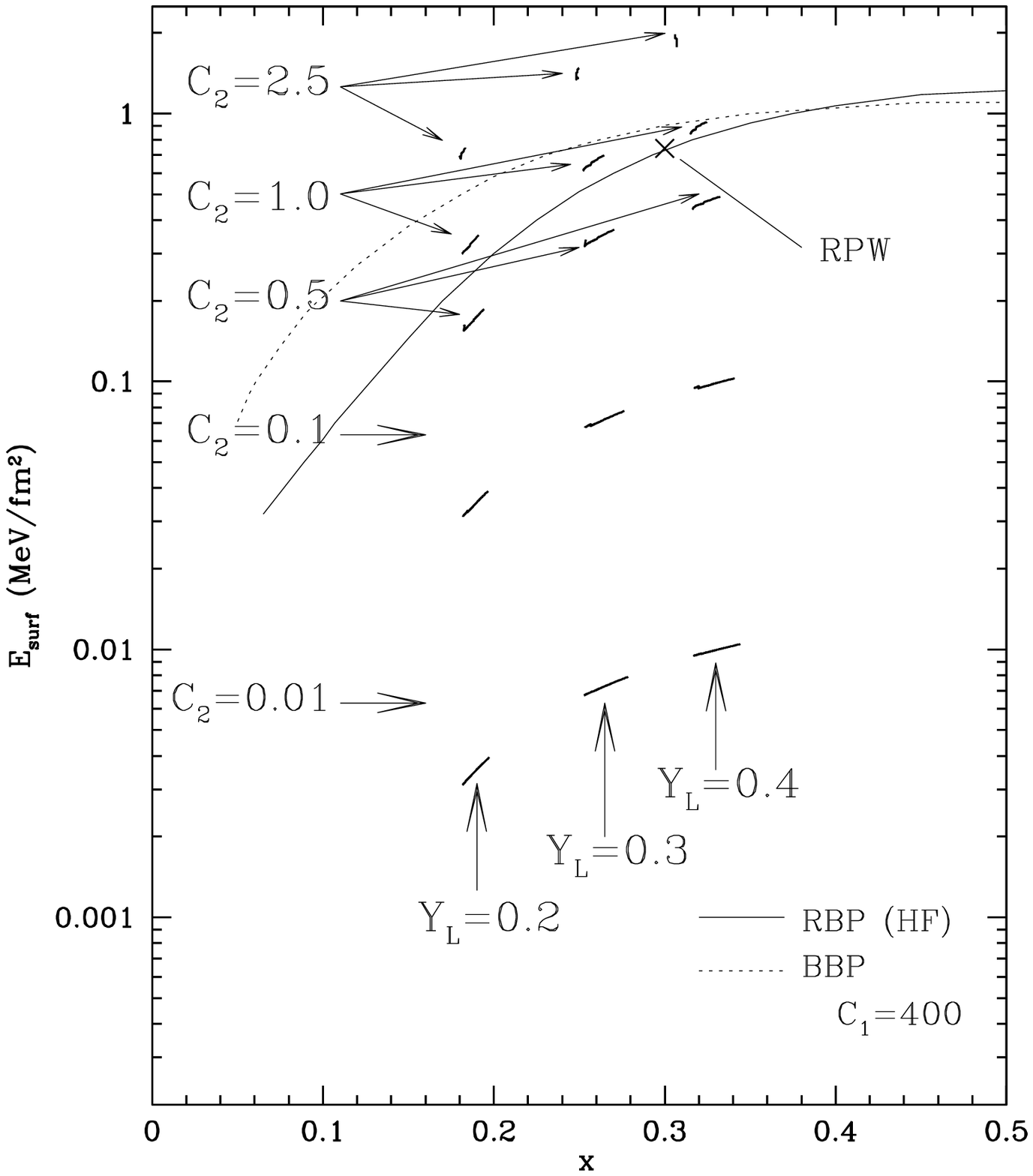}
  \end{center}
\vspace{-20pt}
\caption{}
\label{}
\end{figure}

\newpage
\begin{figure}
  \begin{center}
  \psbox[height=15.5cm]{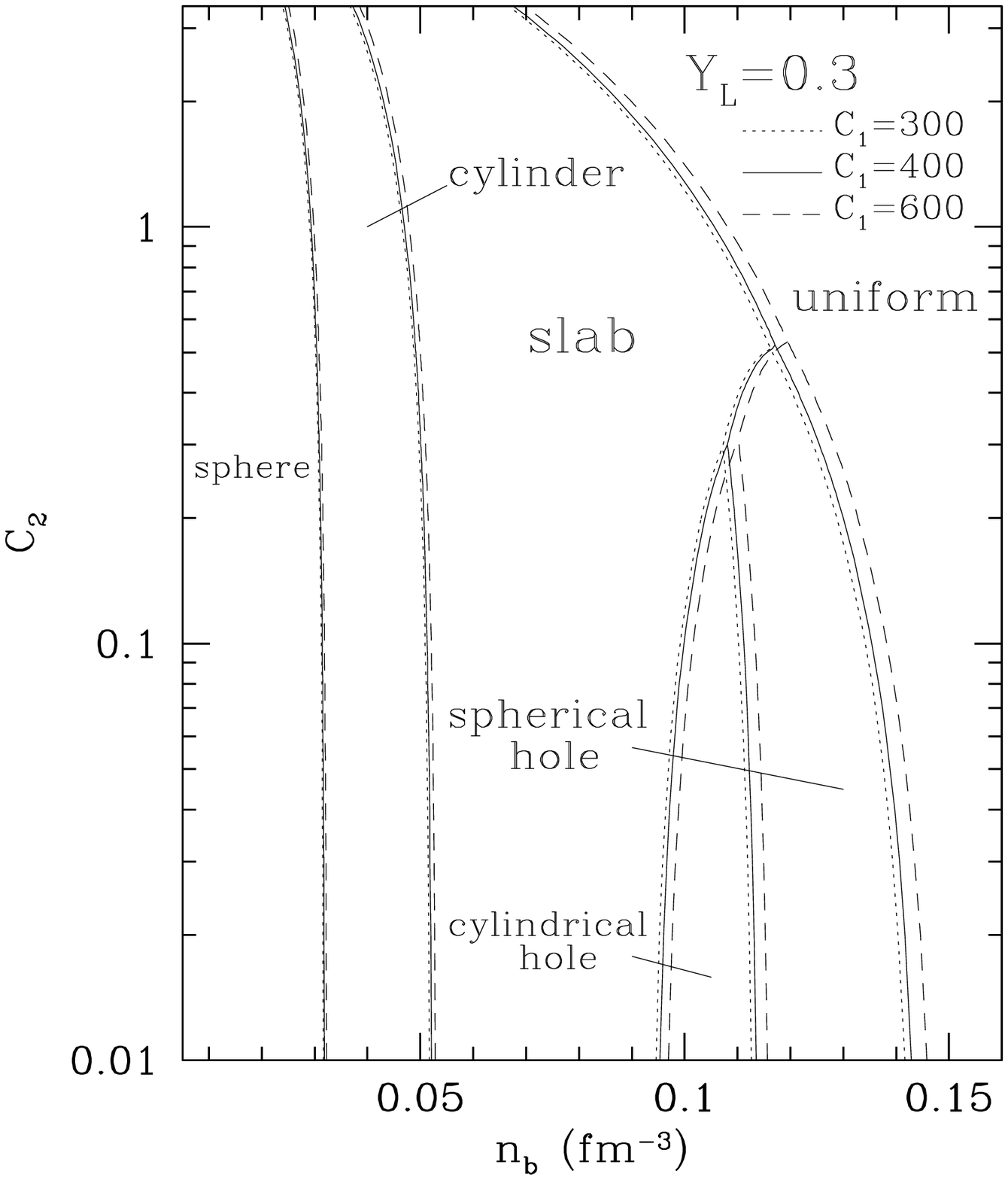}
  \end{center}
\vspace{-20pt}
\caption{}
\label{}
\end{figure}

\newpage
\begin{figure}
\hspace{2cm}
  \psbox[height=18.5cm]{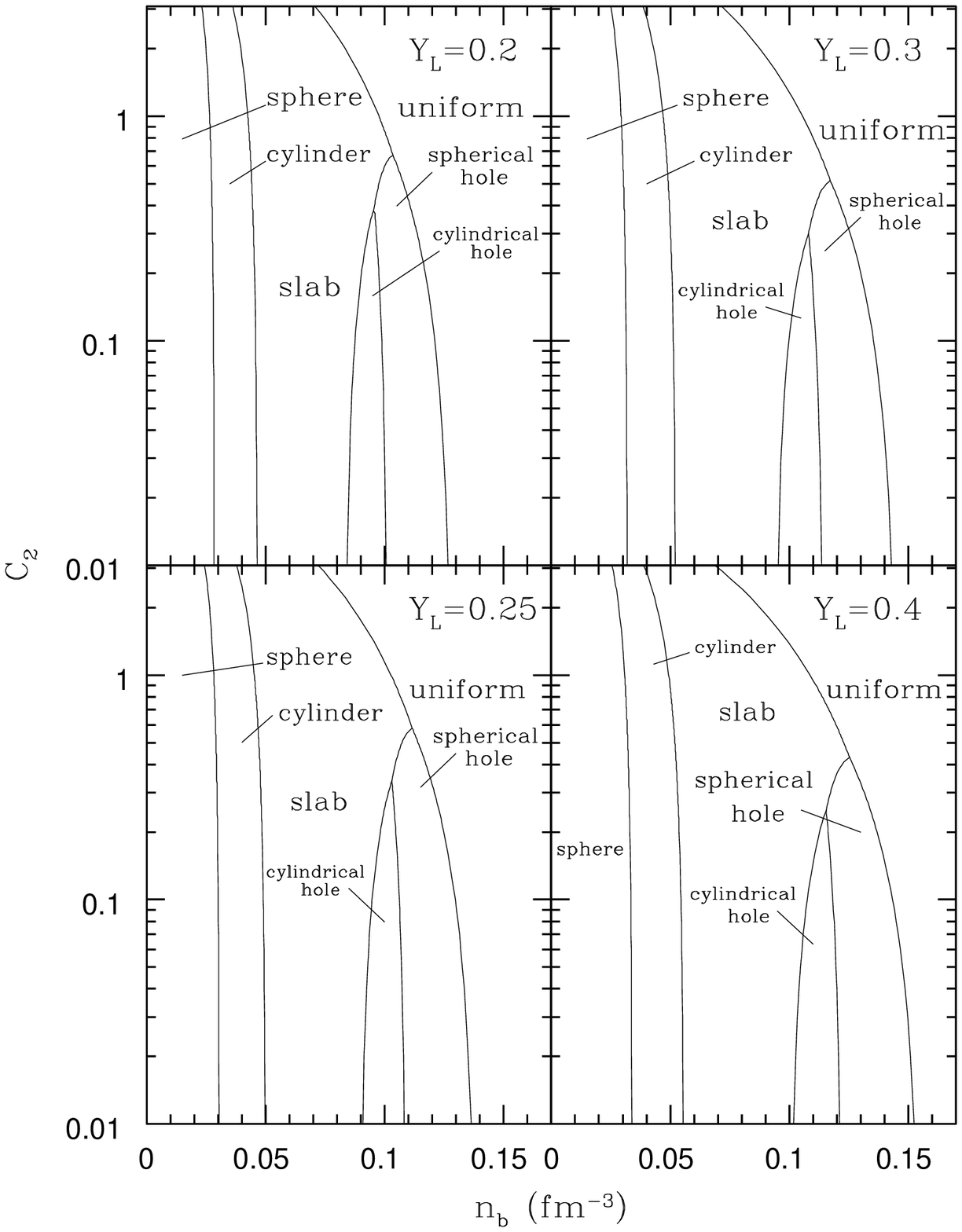}
\vspace{-20pt}
\caption{}
\label{}
\end{figure}

\newpage
\begin{figure}
  \begin{center}
  \psbox[height=18cm]{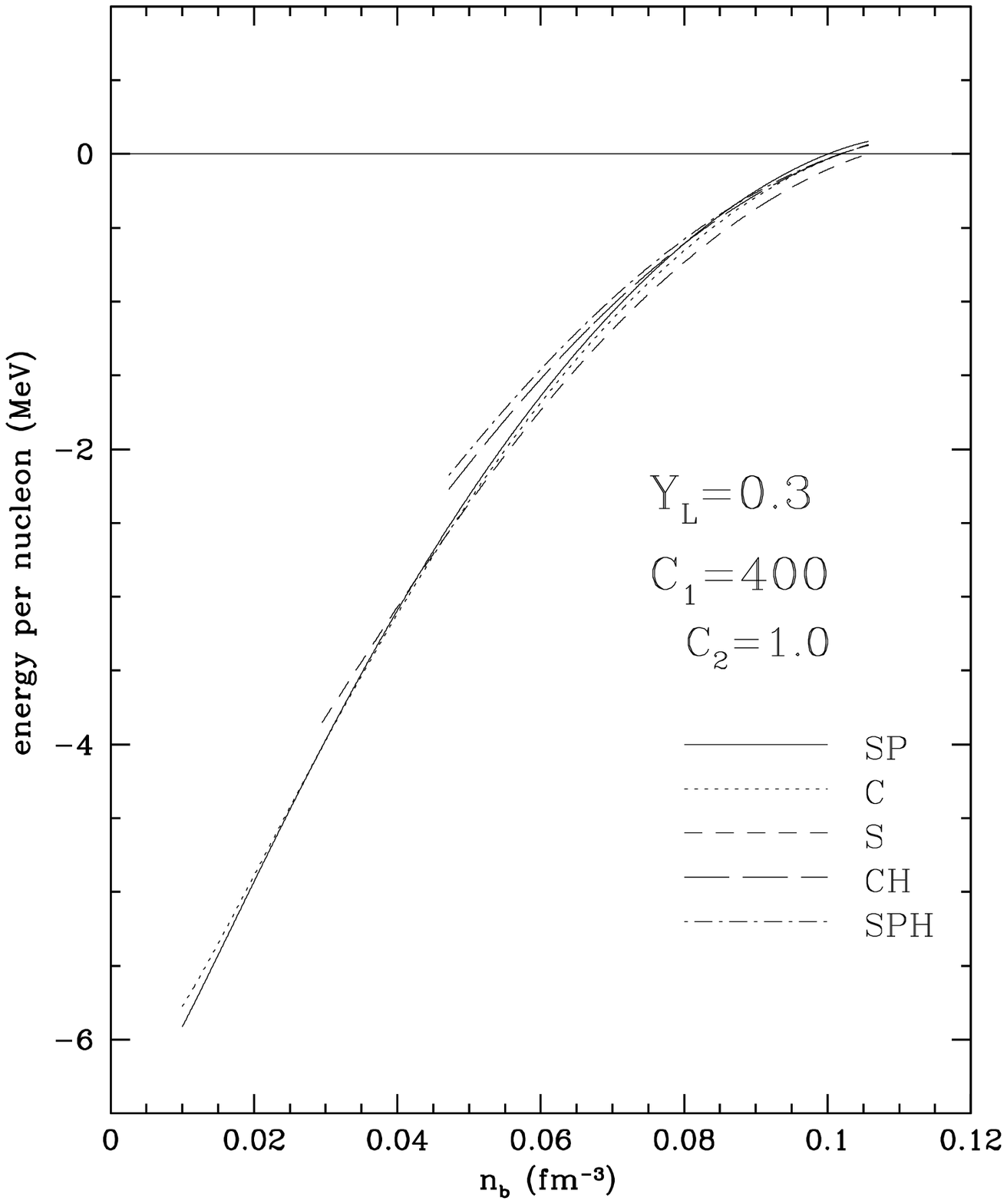}
  \end{center}
\vspace{-20pt}
\caption{}
\label{}
\end{figure}

\newpage
\begin{figure}
  \begin{center}
  \psbox[height=18cm]{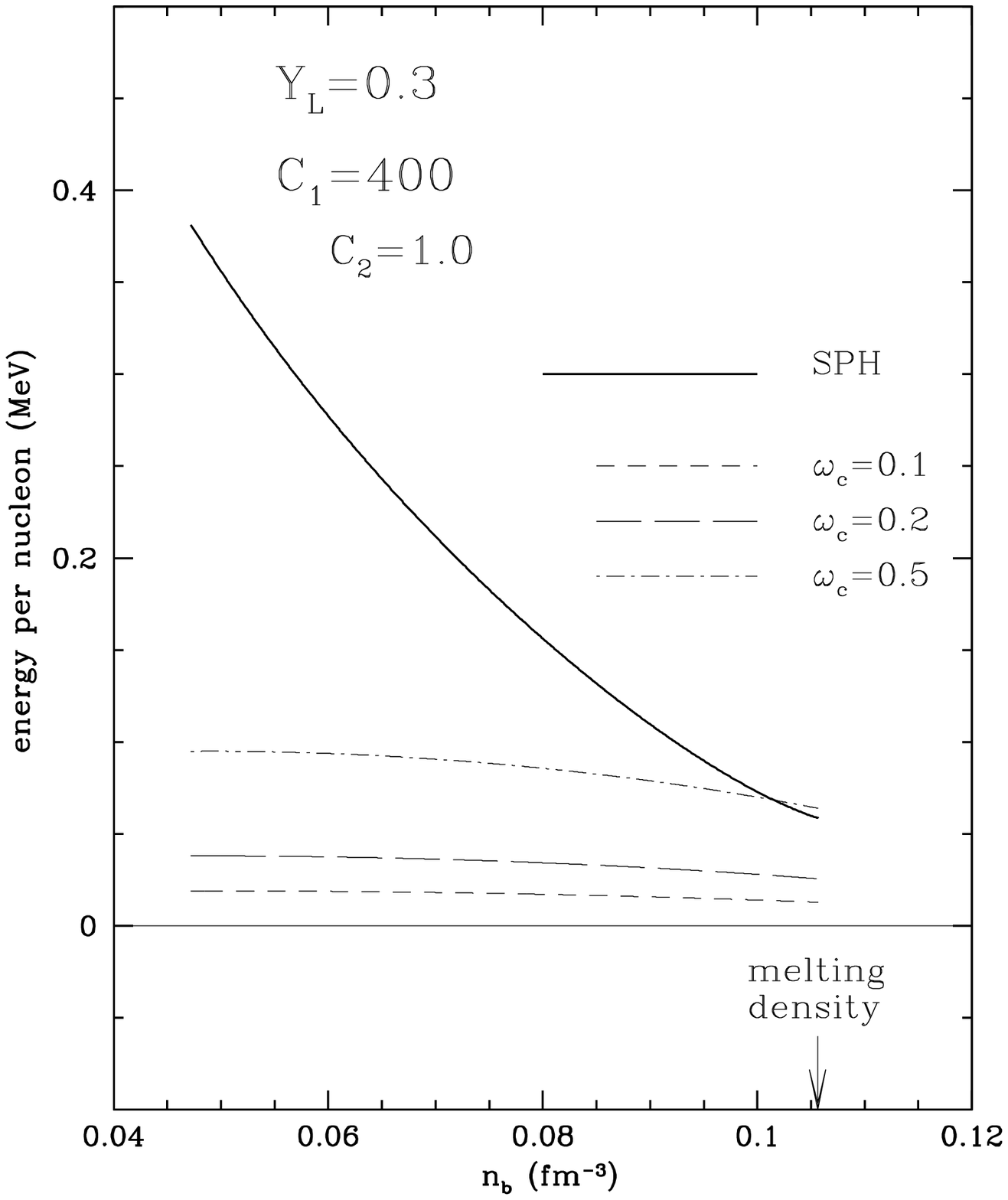}
  \end{center}
\vspace{-20pt}
\caption{}
\label{}
\end{figure}

\newpage
\begin{figure}
  \begin{center}
  \psbox[height=18cm]{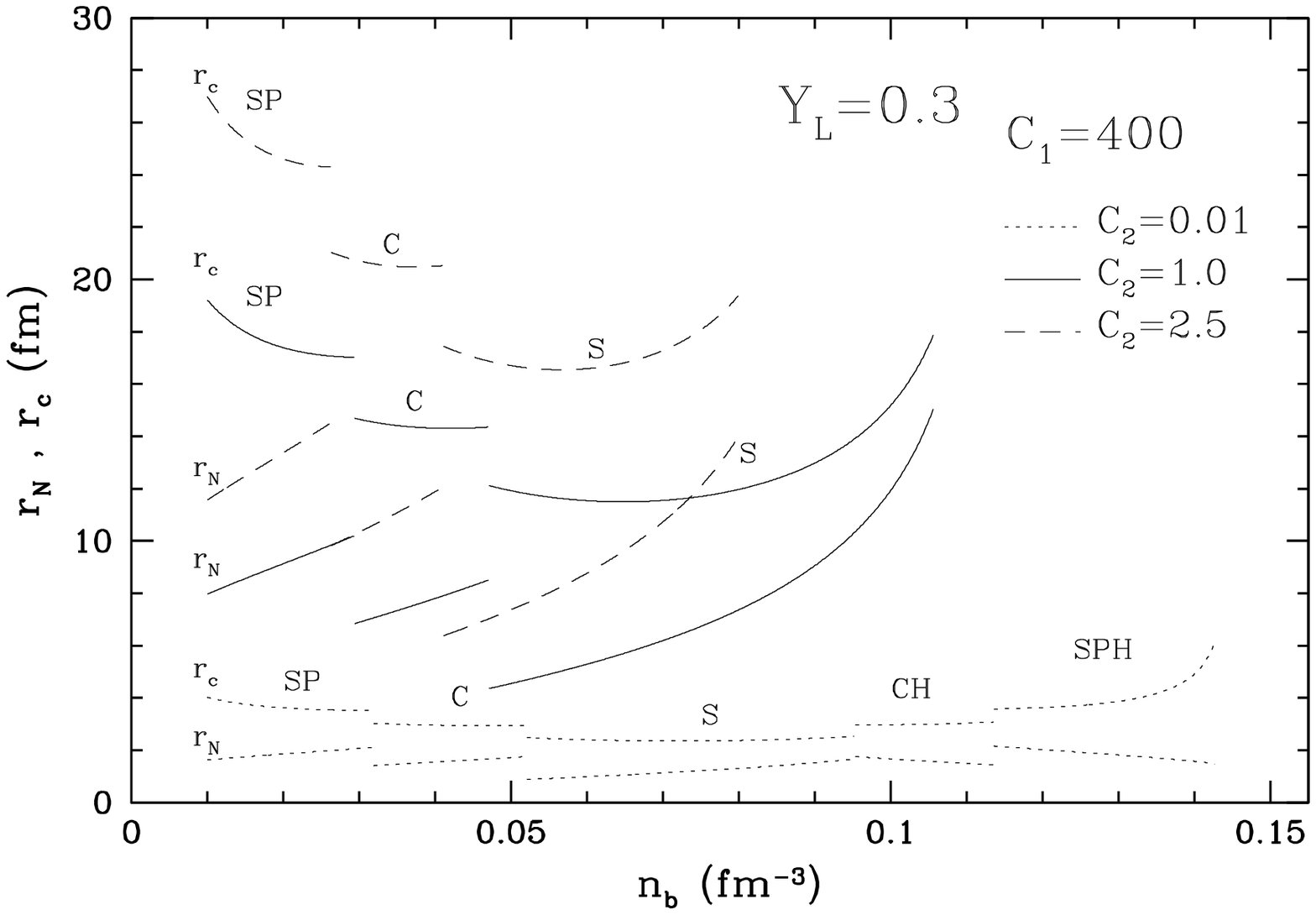}
  \end{center}
\vspace{-20pt}
\caption{}
\label{}
\end{figure}

\newpage
\begin{figure}
  \begin{center}
  \psbox[height=18cm]{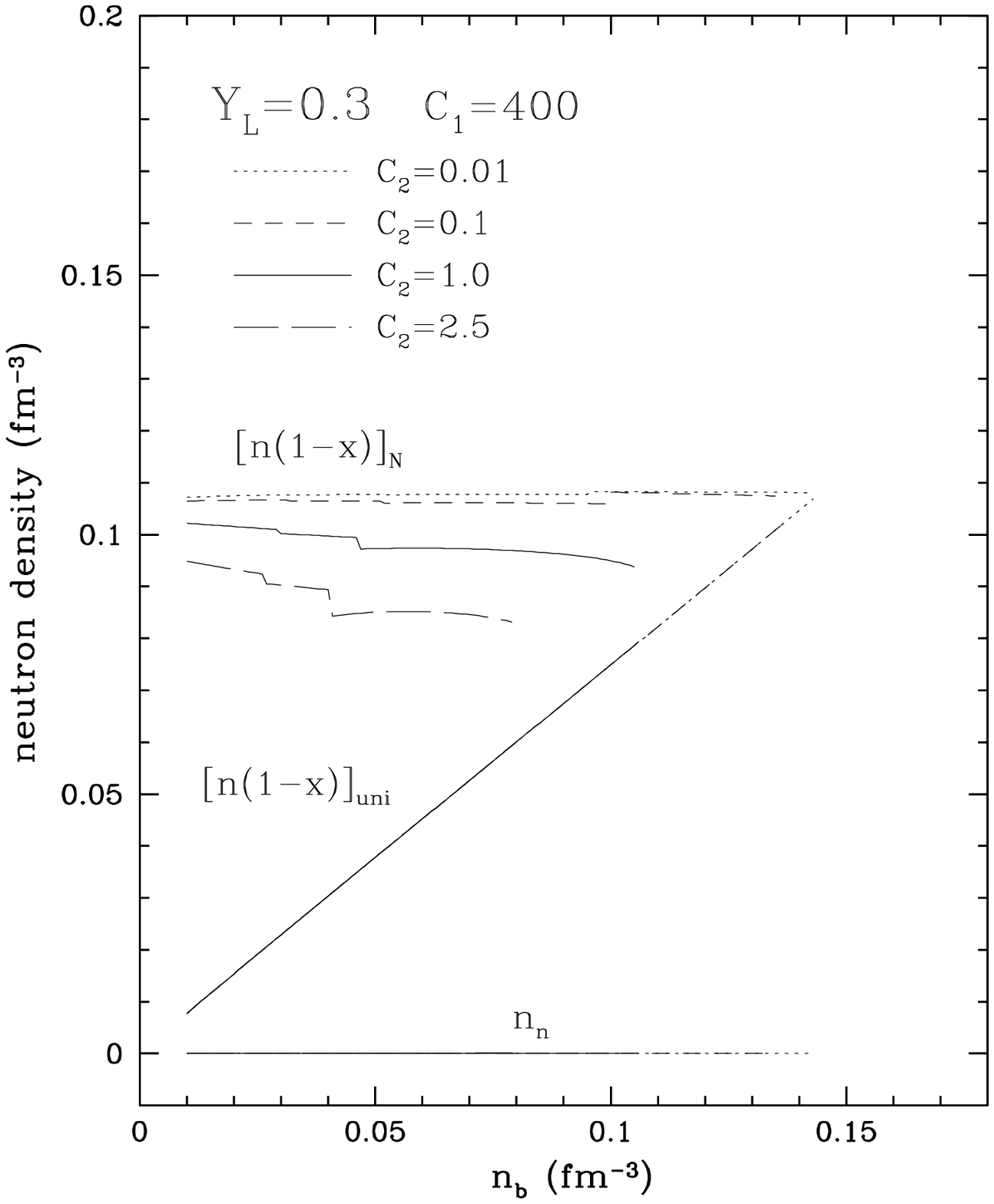}
  \end{center}
\vspace{-20pt}
\caption{}
\label{}
\end{figure}

\newpage
\begin{figure}
  \begin{center}
  \psbox[height=18cm]{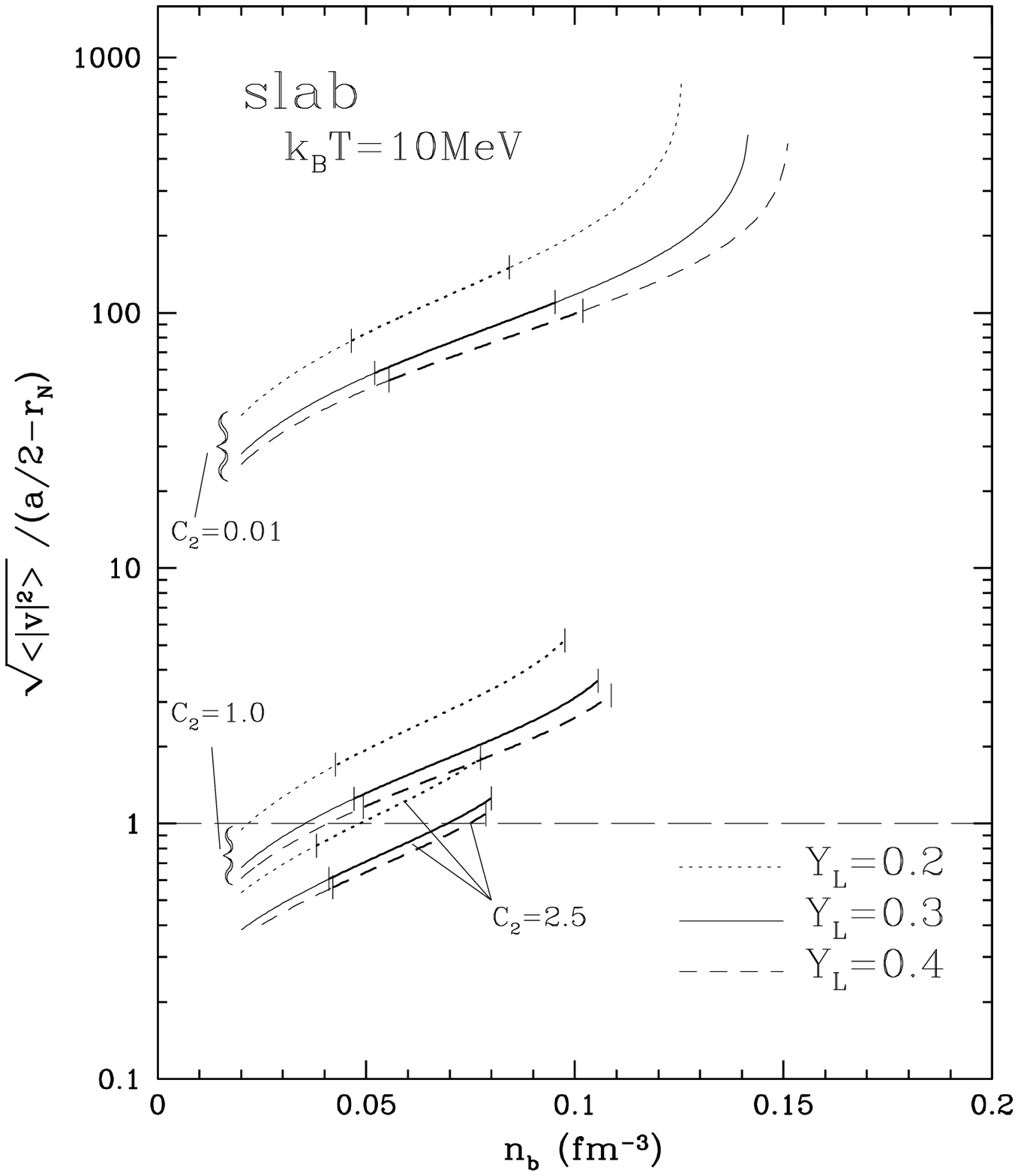}
  \end{center}
\vspace{-20pt}
\caption{}
\label{}
\end{figure}

\newpage
\begin{figure}[p]
  \begin{center}
  \psbox[height=18cm]{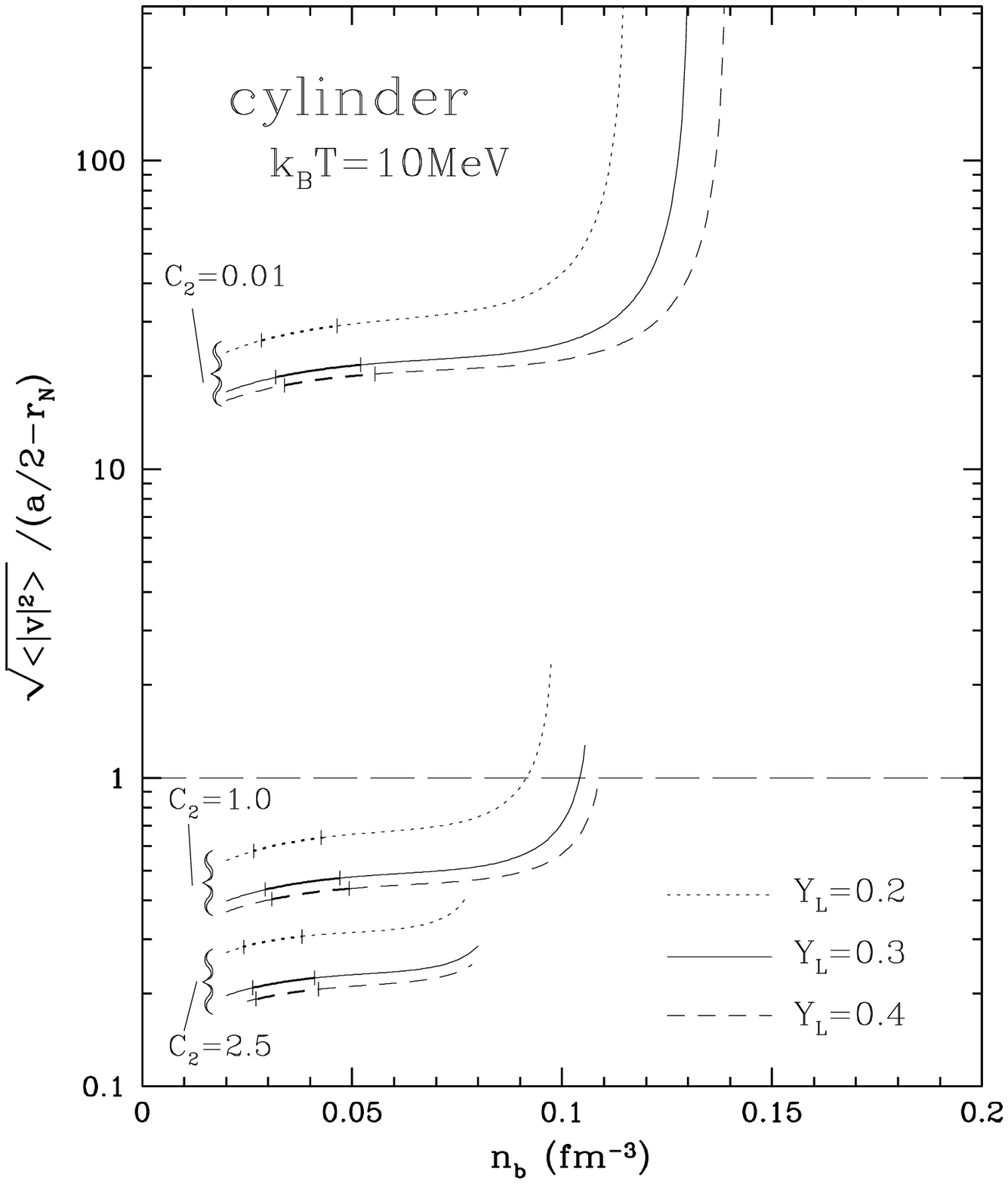}
  \end{center}
\vspace{-20pt}
\caption{}
\label{}
\end{figure}

\newpage
\begin{figure}
  \begin{center}
  \psbox[height=18cm]{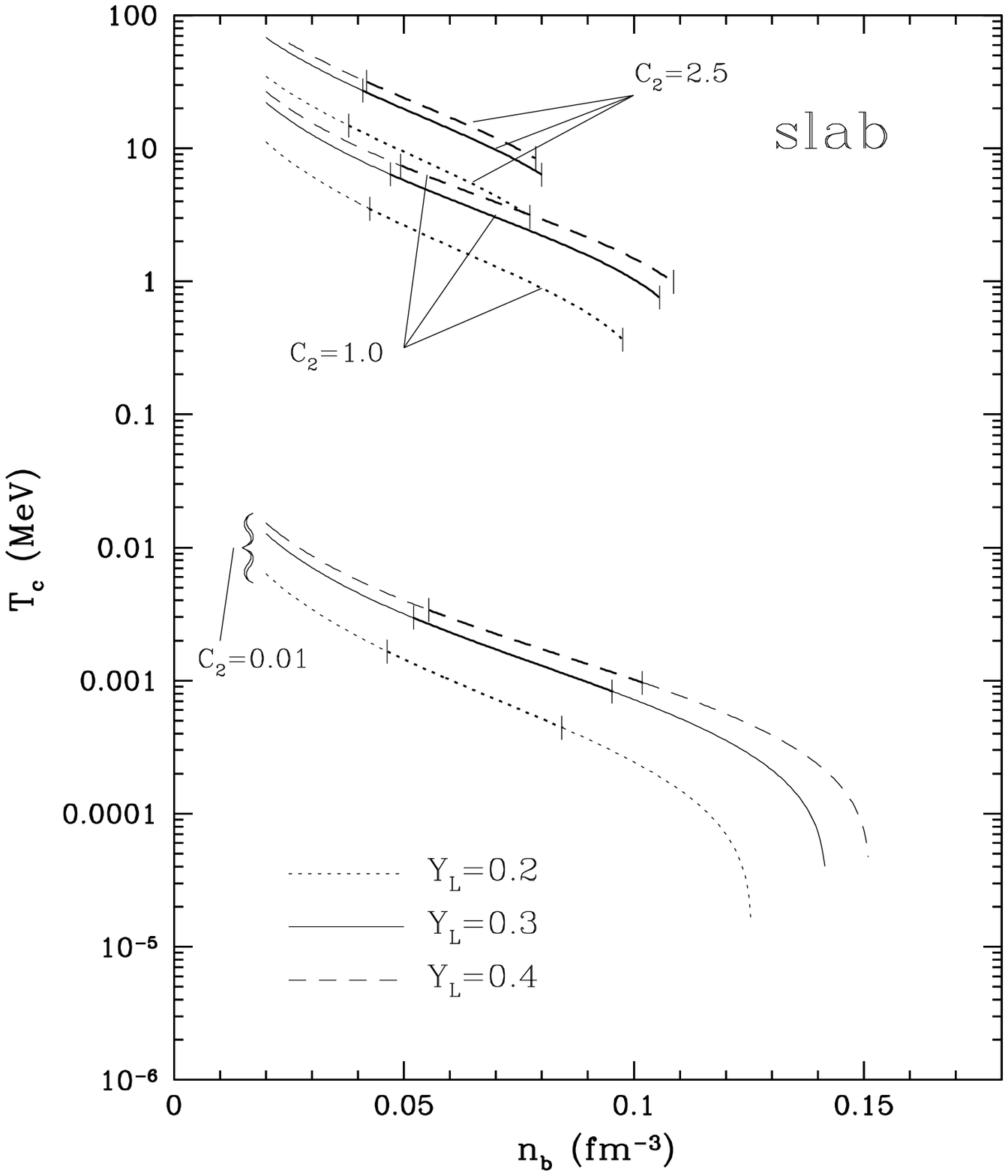}
  \end{center}
\vspace{-20pt}
\caption{}
\label{}
\end{figure}

\newpage
\begin{figure}
  \begin{center}
  \psbox[height=18cm]{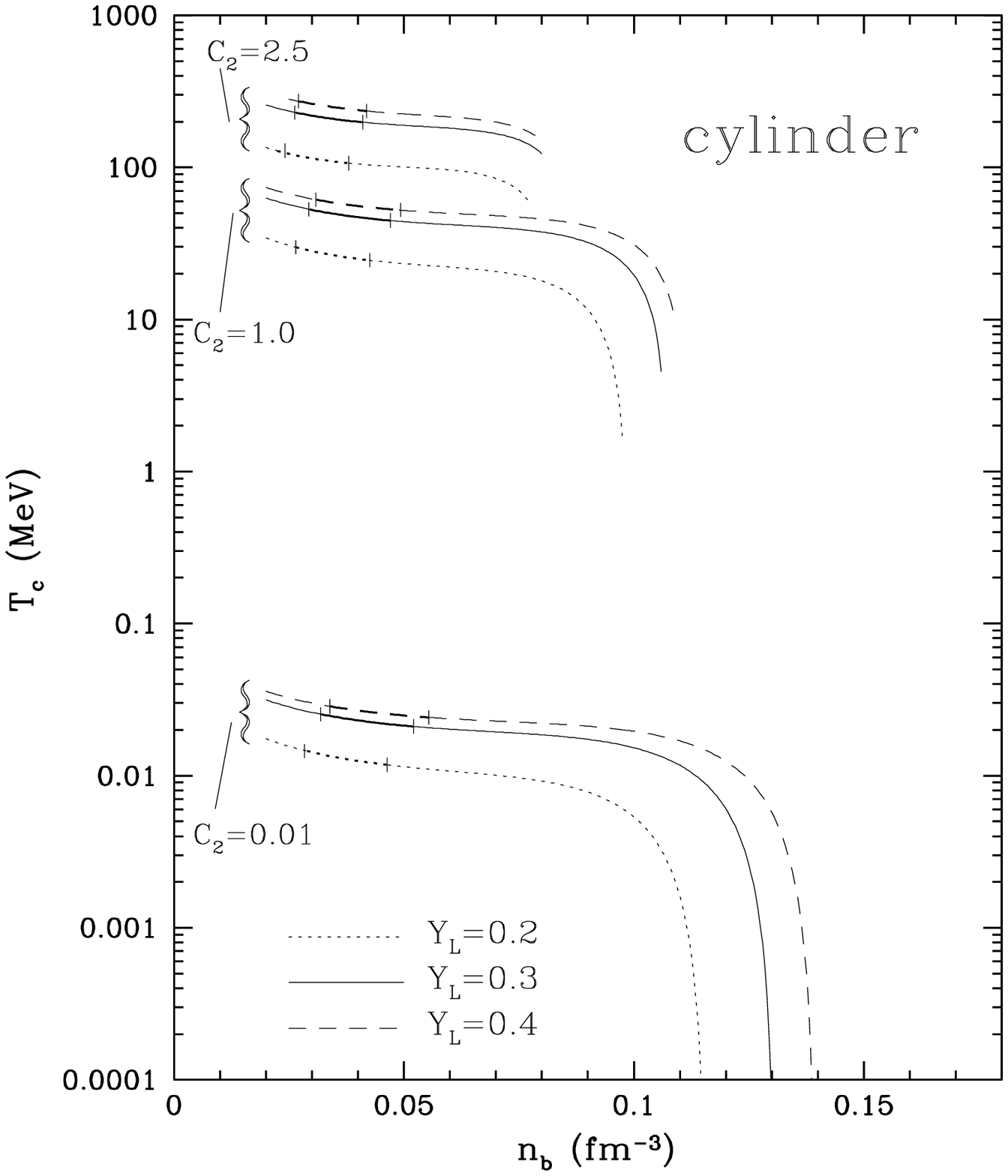}
  \end{center}
\vspace{-20pt}
\caption{}
\label{}
\end{figure}


\newpage

\begin{thebibliography}{99}
%
\bibitem{bethe} H.A. Bethe,
Rev.\ Mod.\ Phys.\ 62 (1990) 801.
%
\bibitem{suzuki} H. Suzuki,
in {\it Physics and Astrophysics of Neutrinos}, 
eds.\ M. Fukugita and A. Suzuki
(Springer, Tokyo, 1994), p.\ 763.
%
\bibitem{freedman} D.Z. Freedman,
Phys.\ Rev.\ D9 (1974) 1389.
%
\bibitem{sato} K. Sato,
Prog.\ Theor.\ Phys.\ 53 (1975) 595; Prog.\ Theor.\ Phys.\ 54 (1975) 1325.
%
\bibitem{ravenhall} D.G. Ravenhall, C.J. Pethick and J.R. Wilson,
Phys.\ Rev.\ Lett.\ 50 (1983) 2066.
%
\bibitem{hashimoto} M. Hashimoto, H. Seki and M. Yamada,
Prog.\ Theor.\ Phys.\ 71 (1984) 320.
%
\bibitem{lassaut} M. Lassaut, H. Flocard, P. Bonche, P.H. Heenen
and E. Suraud,
Astron.\ Astrophys.\ 183 (1987) L3.
%
\bibitem{mezzacappa} A. Mezzacappa, M. Liebend{\"o}rfer,
O.E.B. Messer, W.R. Hix, F.-K. Thielemann and S.W. Bruenn, 
astro-ph/0005366.
%
\bibitem{horowitz} C.J. Horowitz,
Phys.\ Rev.\ D55 (1997) 4577. 
%
\bibitem{bruenn} S.W. Bruenn and A. Mezzacappa, 
Phys.\ Rev.\ D56 (1997) 7529.
%
\bibitem{gentaro} G. Watanabe, K. Iida and K. Sato,
astro-ph/0001273 (to be published in Nucl.\ Phys.\ A). 
%
\bibitem{OTSST} K. Oyamatsu, I. Tanihata, Y. Sugahara, K. Sumiyoshi
and H. Toki, Nucl.\ Phys.\ A634 (1998) 3.
%
\bibitem{BBP} G. Baym, H.A. Bethe and C.J. Pethick,
Nucl.\ Phys.\ A175 (1971) 225.
%
\bibitem{potekhin} C.J. Pethick and A.Y. Potekhin,
Phys.\ Lett.\ B427 (1998) 7.
%
\bibitem{RBP} D.G. Ravenhall, C.D. Bennett and C.J. Pethick,
Phys.\ Rev.\ Lett.\ 28 (1972) 978.
%
\bibitem{arponen} J. Arponen,
Nucl.\ Phys.\ A191 (1972) 257.
%
\bibitem{kolehmainen} K. Kolehmainen, M. Prakash, J.M. Lattimer
and J.R. Treiner,
Nucl.\ Phys.\ A439 (1985) 535.
%
\bibitem{ogasawara} R. Ogasawara and K. Sato,
Prog.\ Theor.\ Phys.\ 68 (1982) 222.
%
\end{thebibliography}
\end{document}